\documentclass[sigconf,nonacm]{acmart}

\usepackage{booktabs}
\usepackage{orcidlink}
\usepackage{tikz}
\usetikzlibrary{arrows.meta,positioning,fit,shapes.geometric,calc}

\tikzset{
  box/.style   = {draw, rounded corners=2pt, align=center, font=\scriptsize,
                  minimum height=7mm, inner xsep=4pt},
  spec/.style  = {box, fill=blue!8},
  ver/.style   = {box, fill=green!10},
  ctrl/.style  = {box, fill=gray!12},
  exec/.style  = {box, fill=orange!12},
  mem/.style   = {box, fill=red!8},
  gate/.style  = {draw, diamond, aspect=2.2, align=center, font=\scriptsize,
                  fill=green!10, inner sep=1pt},
  flow/.style  = {-{Latex[length=2mm]}, thick},
  back/.style  = {-{Latex[length=2mm]}, thick, dashed},
  pair/.style  = {dotted, thick, gray},
  mediate/.style = {dotted, thick, gray,
                    {Latex[length=1.5mm]}-{Latex[length=1.5mm]}},
}

\makeatletter
\def\@secfont{\normalfont\Large\scshape\section@raggedright}
\def\@subsecfont{\normalfont\large\itshape\section@raggedright}
\def\@subsubsecfont{\normalfont\normalsize\itshape}
\makeatother

\setcopyright{none}
\renewcommand\footnotetextcopyrightpermission[1]{}

\begin{document}

\title{VReAct: Verification as an Architectural Layer for LLM Agents}
\subtitle{A V-Model Design, and a Pilot Study of Its Deterministic Core}

\author{Ali Afoud}
\orcid{0009-0006-7204-993X}
\affiliation{%
  \institution{Independent Researcher}   
  \city{}                                
  \country{Canada}
}

\author{Jie JW Wu}
\orcid{0000-0002-7895-2023}
\affiliation{%
  \institution{Michigan Technological University}
  \city{Houghton}
  \country{USA}
}

\begin{abstract}
Large language model (LLM) agents built on the ReAct pattern concentrate four
responsibilities in a single model: selecting a strategy, choosing each action,
formatting it, and judging whether the result is adequate. No component outside
the generative loop can reject the loop's output, so an agent that cannot make
progress does not report failure; it continues until an externally imposed
budget terminates it. We propose VReAct, where we treat verification as an architectural layer
for ReAct agent systems by adapting the V-model from software engineering:
specification levels descend from requirements to individual steps, each level
is paired with a dedicated verifier, a deterministic controller enforces every
verdict, and memory is written only by verification outcomes, so that a
rejection localizes the level that introduced the fault and an agent halts by
declining rather than by exhaustion. Each verifier separates a zero-cost
deterministic \emph{gate} from an optional LLM \emph{judge}, allowing the
contribution and cost of each to be measured independently. We report a pilot
study implementing the acceptance-level and unit-level verifier pairs, comparing
five configurations that share one executor, tool set, and scorer and differ
only in their verification components, on the four-hop stratum of MuSiQue with
an 8B-parameter backbone. Across 47 executions, the two unverified
configurations answered none of ten questions, every run ending at a step cap or
provider token limit; the verified configuration without a planner answered
eight of ten and abstained on the rest. Deterministic gates produced eight of
the nine observed corrections at zero marginal cost, and planning degraded
performance once verification was present. These results characterize
termination behavior rather than accuracy at scale; we outline a twelve-month
plan for completing and evaluating the full architecture, including the
integration-level pair the pilot omits.
\end{abstract}

\maketitle
\section{Introduction}

An LLM agent is a language model in a control loop: it receives a goal, invokes
tools, and continues until it decides it has finished. The prevailing design is
ReAct~\cite{react}, which extends chain-of-thought prompting~\cite{cot} to an
interactive setting by alternating a \emph{Thought}, which modifies only the
model's context, with an \emph{Action}, whose \emph{Observation} is appended to
that context. Most deployed frameworks are this loop with a tool registry
attached~\cite{autogen,toolformer}, and the pattern is now common in software
engineering~\cite{hou,fanllm4se}. The design concentrates four responsibilities
in one model and one prompt: selecting the overall strategy, selecting each
action, formatting that action, and judging whether the resulting observation is
sufficient. No component outside the loop can reject the loop's output.

The practical problem is not primarily accuracy but the absence of a
\emph{mechanism for declining}. A ReAct loop that cannot make progress does not
report failure; it continues to emit actions until an externally imposed budget
intervenes or, absent a budget, until the serving infrastructure refuses further
requests. Such an agent cannot be composed into a larger system, because a
caller receives either a plausible answer or nothing and cannot distinguish the
two. The literature documents the underlying mechanics: turn-by-turn action
selection is a greedy policy that can be arbitrarily suboptimal over long
horizons, with reported myopic-trap rates above 50\% and post-error recovery
near 5\%~\cite{flare}; agents abandon plans they produced, repeat executed
steps, and act inconsistently with the preceding thought~\cite{eog}; and token
consumption grows quadratically because the full history is resubmitted each
turn~\cite{rewoo}. A larger budget does not help: the agent commits to an
unproductive trajectory early~\cite{rpreact}.

Existing work addresses the symptoms rather than the structure. Planning and
efficiency methods change what the agent reads or how far ahead it
looks~\cite{rewoo,rpreact,flare,adapt}, and prompting strategies improve how it
reasons~\cite{selfconsistency,tot}, but neither inspects the result. Where
checking does exist, it is the agent checking itself~\cite{reflexion,selfrefine},
which is the configuration that Huang et al.~\cite{selfcorrect} report to fail
and that Zheng et al.~\cite{mtbench} report to be biased. Table~\ref{tab:gaps}
summarizes the four gaps that remain and how this work addresses each; the
pilot closes G3 and makes partial progress on two others.

\begin{table*}[t]
\centering
\small
\setlength{\tabcolsep}{6pt}
\caption{The four gaps this work addresses.}
\label{tab:gaps}
\begin{tabular}{@{}p{0.045\textwidth}p{0.28\textwidth}p{0.28\textwidth}p{0.31\textwidth}@{}}
\toprule
& \textbf{What the literature does} & \textbf{The gap} & \textbf{Our fix} \\
\midrule
\textbf{G1} &
The model that produced the output also checks it~\cite{reflexion,selfrefine}. &
No agent has a checker \emph{outside} the loop whose verdict it must obey. Self-checking
without external feedback degrades accuracy~\cite{selfcorrect,mtbench}. &
Verifiers are separate components. The controller enforces their verdicts, so the
agent cannot proceed past a rejection. \\
\addlinespace
\textbf{G2} &
External checks exist, but for one artifact type: a reasoning
step~\cite{verifystep}, a program~\cite{selfdebug}. &
None pairs checks to the levels of a specification, so a rejection cannot say
\emph{which} stage broke, and recovery means starting over. &
One verifier per level. A rejection names the level that failed, and only that
level is redone. \\
\addlinespace
\textbf{G3} &
Systems mix free structural checks with LLM judgement and report the combined
effect. &
Nobody has measured what each half contributes, so the layer cannot be priced. &
Gates and judges are separate stages, and every correction is attributed to one of
them. \\
\addlinespace
\textbf{G4} &
Context is trimmed to fit a budget~\cite{rewoo,rpreact}. &
Nothing decides what deserves to be carried forward: verified and failed work are
resubmitted alike. &
Only verdicts write memory. Accepted results go to a ledger, rejections to an index
keyed by component and cause. \\
\bottomrule
\end{tabular}
\end{table*}

Our response is architectural: we argue that the missing structure is a
verification layer, and that software engineering already provides its template.
Software engineering separated construction from certification decades ago, and
the V-model pairs each design stage with a test stage of matching scope. We
transfer that structure to LLM agents. It is the architecture, rather than the
pilot study, that this line of work ultimately seeks to establish.

\subsection{What ``verification'' means here}
\label{sec:defn}

The term \emph{verification} is used loosely in the agent literature, covering
self-critique, self-consistency, confidence estimation, reranking, and formal
proof; in LLM training it now also names reward models scoring reasoning
steps~\cite{verifystep}. We use it in neither sense but in the
software-engineering one~\cite{boehm}, and we claim no formal guarantees: our
checks decide predicates on one artifact of one execution, as in runtime
verification~\cite{runtimever}, not properties over all executions. A
\emph{verification act} satisfies four conditions.
\begin{itemize}\setlength{\itemsep}{1pt}
\item \emph{V1, externality}: the verdict comes from a component distinct from
  the one that produced the artifact: a restricted-input invocation, ordinary
  program logic, or, preferably, a separate model.
\item \emph{V2, authority}: the verdict is binding and enforced by the
  controller; it is not returned as advice the generator may ignore.
\item \emph{V3, pairing}: each verifier inspects artifacts at exactly one
  specification level, so the level at which a rejection fires identifies where
  the fault was introduced and scopes the retry.
\item \emph{V4, bounded termination}: each level's rejection budget is finite,
  and exhausting it yields an abstention naming the level that failed, not a
  degraded answer.
\end{itemize}

Three terms recur. A \emph{gate} is a deterministic verifier written as program
logic---a regular expression, a set-membership test, or, in code, a compiler or
test runner; it costs nothing per invocation and is self-consistent, though
determinism buys repeatability, not correctness: a gate passes any artifact
wrong in a way its predicate does not express (\S\ref{sec:qaexample}). A
\emph{judge} is a model invocation posed one question that no gate can answer;
its authority derives from the controller that enforces its verdict, not from
any assumed reliability of the model (\S\ref{sec:selfassess}). A \emph{guard}
is loop control rather than verification: it decides whether an action is
re-executed, whether unparseable output ends the run, and how much of an
observation enters the transcript, but it evaluates nothing for correctness.
The third term matters because guards and gates are easily conflated and, in
the pilot, confounded (\S\ref{sec:confound}).

\subsection{Objectives}
\label{sec:objectives}

We pursue two objectives of different scope, stated separately because the
evidence for each differs.

\textbf{O1, near-term: a minimal verification layer is a cheap and
near-universal improvement to ordinary agents on ordinary tasks.} A small set of
deterministic checks---costing no model calls and expressible in a few hundred
lines of ordinary code---changes the behavior of a conventional ReAct agent out
of proportion to its cost: primarily by making the agent \emph{stop for a stated
reason}, and secondarily by making it more accurate. The claim's scope is deliberately modest---short tasks, one model
tier, one task family---and if it holds, the recommendation is unconditional,
because the layer is free at run time and does not depend on model capability.

\textbf{O2, long-term: verification should be a standard architectural
commitment for agent systems.} In software engineering, a test layer is not an
optimization adopted when time permits; it is structural, and its absence is a
defect in the process. We argue that agent systems are in that state now.
Concretely, an agent built to this proposal returns exactly one of two things:
an answer that named verifiers accepted, or an abstention naming the level that
refused and the evidence it saw---never an answer no component but its author
has inspected, and never a stop forced by a token limit. The V-model pairing is
what provides these guarantees: specification levels, a verifier paired with
each level, deterministic control of every transition, and memory written only
by verdicts. The target regime is the one in which agents are simultaneously
least reliable and most in demand: long-horizon tasks with many interdependent
components, where one unchecked intermediate result contaminates everything
downstream. We do not claim that this paper establishes O2. We claim that O2 is
well posed, that it is a candidate for a standard rather than one more
technique, and that this paper provides its first measurement.

Three observations make O2 plausible rather than merely appealing. First, most
of the layer is \emph{free}: the deterministic half produced eight of the nine
corrections in our study at zero marginal cost. Second, the layer does
\emph{not require a stronger model}: the gates worked at the 8B scale, at which
the executor itself barely worked. Third, the structure is already validated in
the discipline it is borrowed from, and the evidence against
self-correction~\cite{selfcorrect,mtbench} argues \emph{for} externalizing the
check rather than against checking. The counterweight is our own planning
result (\S\ref{sec:plan}): adding a decomposition stage without the verifier
meant to check it made the system markedly worse, so the structure may be
adoptable only as a whole---a cost the plan in \S\ref{sec:roadmap} must assess.

\subsection{The pilot, and its scope}
\label{sec:contrib}

The pilot study implements the outer (L1/R3) and inner (L3/R1) pairs of the
architecture and compares five configurations that differ by one component at a
time, so that every correction can be attributed to the component that caused
it. Its research questions test O1 directly and bound O2 from below.
\textbf{RQ1}: What does a deterministic verification layer do to task success
and to termination behavior? \textbf{RQ2}: What does an LLM judge add beyond
the gates? \textbf{RQ3}: Does planning still help once verification is present?
\textbf{RQ4}: What does each component cost?

We designed the study around \emph{silent errors}: answers that are wrong but
delivered without any indication that the chain failed. At the model scale we
study, that failure mode did not occur---the unverified configurations never
advanced far enough to be confidently wrong, and instead collapsed structurally.
The contribution therefore shifted to \emph{termination behavior}: whether an
agent halts for a stated reason or continues until refused externally, a
property that matters comparably for deployment. The silent-error hypothesis is
reported as untested at this scale, with qualitative observations at larger
scale (\S\ref{sec:stronger}). The evidence base is 47
executions over five items with one backbone, so all quantitative statements are
directional; the middle pair (\S\ref{sec:middle}) and the memory schema
(\S\ref{sec:memory}) are specified but not built.

In summary, this paper makes the following contributions:
\begin{itemize}\setlength{\itemsep}{1pt}
\item an architecture that treats verification as a layer of an agent system,
  pairing each specification level with a verifier under a deterministic
  controller (\S\ref{sec:vision});
\item a definition of verification for agents (V1--V4) separating gates,
  judges, and guards, which makes each half of the layer separately measurable
  (\S\ref{sec:defn});
\item a pilot study of the deterministic core: five configurations, 47
  executions, every correction attributed to its component
  (\S\ref{sec:results});
\item a twelve-month plan that supplies what the pilot lacks and moves to a
  code domain where the integration pair can be exercised
  (\S\ref{sec:roadmap}).
\end{itemize}

\paragraph{Workshop positioning, and the feedback we seek.}
This paper is a pilot, submitted to JAWs for feedback on experimental design
before the larger study is committed to, and we would value disagreement on
three points: (i)~gates and guards are confounded (\S\ref{sec:confound})---is
separating them the right next controlled condition, or should scaling across
model tiers come first? (ii)~the judges are the same model that produced the
work, and two repairs are available (\S\ref{sec:selfassess})---a stronger judge
above a weaker executor, or a judge restricted to questions the deterministic
layer can ground: which makes the better second study? (iii)~the integration
verifier needs a task family with assembled artifacts, for which we propose
SWE-bench---would a smaller domain with cheaper ground truth serve better?

\section{Background}
\label{sec:background}

Verification and validation are distinct activities with distinct
artifacts~\cite{boehm}. In the V-model, each design stage is paired with a test
stage of matching scope: unit tests examine the detailed design, integration
tests examine the architecture, and acceptance tests examine the requirements.
Falc\~ao et al.~\cite{vmodelresearch} report nearly two decades of its use as a
framing device whose stage transitions act as quality gates, and
Wu~\cite{vmodelml} finds, from practitioner interviews on ML-enabled systems,
that layered decomposition, explicit responsibility boundaries, and consistent
verification across levels correspond to those systems' failure modes. Our proposal rests on the observation that an agent's control
loop is a delivery process of this kind---run at inference time and at high
frequency---but built without the corresponding certification structure.

The failure mode that model-based verification targets, a fluent, confident, and
unsupported answer, is well characterized~\cite{hallucination}; detection
methods range from sampling-based consistency~\cite{selfcheckgpt} to attribution
against context spans~\cite{attntrace}, and our deterministic grounding check is
the simplest member of that family, testing whether the answer occurs in any
returned observation. For the task domain, MuSiQue~\cite{musique} was
constructed so that single-hop shortcuts fail, which is why we adopt it over
HotpotQA~\cite{hotpotqa}; agentic approaches to multi-hop retrieval include
Self-Ask~\cite{selfask} and IRCoT~\cite{ircot}. Broader agent
benchmarks~\cite{agentbench,taubench} report that long-horizon tasks fail for
structural reasons and that single-attempt accuracy overstates deployed
reliability.

\section{The Proposed Architecture: ReAct under the V-Model}
\label{sec:vision}

\subsection{Levels and their paired verifiers}

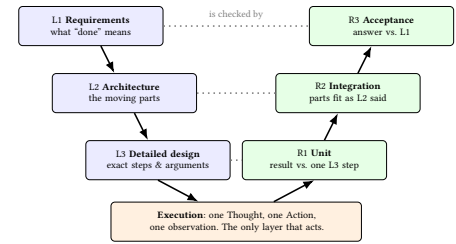
\begin{figure}[b]
\centering
\resizebox{0.68\columnwidth}{!}{%
\begin{tikzpicture}[x=1mm,y=1mm,
      spec/.append style={text width=23mm, font=\tiny},
      ver/.append style ={text width=23mm, font=\tiny}]
  \node[spec] (l1) at (0,0)    {L1 \textbf{Requirements}\\what ``done'' means};
  \node[spec] (l2) at (6,-12)  {L2 \textbf{Architecture}\\the moving parts};
  \node[spec] (l3) at (12,-24) {L3 \textbf{Detailed design}\\exact steps \& arguments};

  \node[ver]  (r3) at (52,0)   {R3 \textbf{Acceptance}\\answer vs.\ L1};
  \node[ver]  (r2) at (46,-12) {R2 \textbf{Integration}\\parts fit as L2 said};
  \node[ver]  (r1) at (40,-24) {R1 \textbf{Unit}\\result vs.\ one L3 step};

  \node[exec, text width=42mm, font=\tiny] (ex) at (26,-35)
    {\textbf{Execution}: one Thought, one Action,\\one observation.
     The only layer that acts.};

  \draw[flow] (l1) -- (l2);   \draw[flow] (l2) -- (l3);
  \draw[flow] (l3.south) -- (ex);
  \draw[flow] (ex) -- (r1.south);
  \draw[flow] (r1) -- (r2);   \draw[flow] (r2) -- (r3);
  \draw[pair] (l1) -- node[above,font=\tiny,gray]{is checked by} (r3);
  \draw[pair] (l2) -- (r2);   \draw[pair] (l3) -- (r1);
\end{tikzpicture}}
\caption{The motivating frame. Solid arrows show the flow of work; dotted lines
show which verifier checks which specification. The pilot implements the outer
pair (L1/R3) and the inner pair (L3/R1) only.}
\Description{A V-shaped diagram. Three specification levels descend on the left, requirements to architecture to detailed design, meeting an execution box at the bottom; three verification levels ascend on the right, unit to integration to acceptance. Dotted horizontal lines pair each specification level with the verifier of matching scope.}
\label{fig:vmodel}
\end{figure}

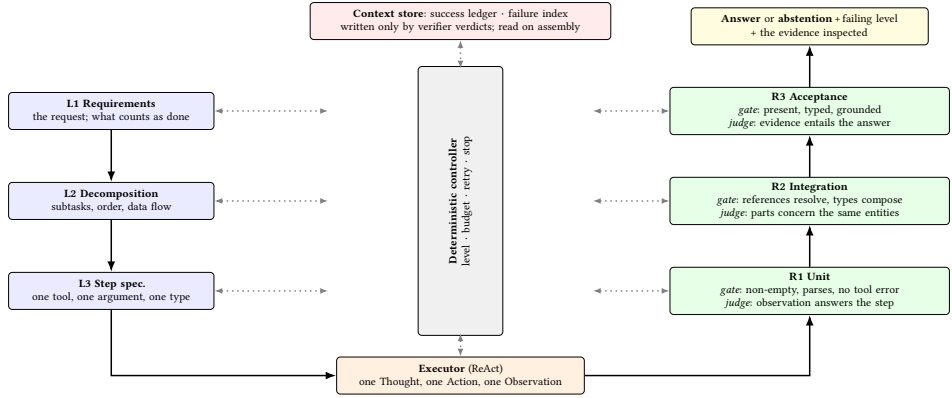
\begin{figure*}[t]
\centering
\resizebox{0.70\textwidth}{!}{%
\begin{tikzpicture}[x=1mm,y=1mm,
   spec/.append style={text width=36mm, font=\scriptsize},
   ver/.append style={text width=50mm, font=\scriptsize}]

  \node[spec] (l1) at (0,0)    {\textbf{L1 Requirements}\\ the request; what counts as done};
  \node[spec] (l2) at (0,-17)  {\textbf{L2 Decomposition}\\ subtasks, order, data flow};
  \node[spec] (l3) at (0,-34)  {\textbf{L3 Step spec.}\\ one tool, one argument, one type};

  \node[ver] (r3) at (132,0)   {\textbf{R3 Acceptance}\\ \emph{gate}: present, typed, grounded\\ \emph{judge}: evidence entails the answer};
  \node[ver] (r2) at (132,-17) {\textbf{R2 Integration}\\ \emph{gate}: references resolve, types compose\\ \emph{judge}: parts concern the same entities};
  \node[ver] (r1) at (132,-34) {\textbf{R1 Unit}\\ \emph{gate}: non-empty, parses, no tool error\\ \emph{judge}: observation answers the step};

  \node[ctrl, rotate=90, minimum width=50mm, minimum height=16mm,
        text width=48mm, font=\scriptsize] (c) at (66,-17)
    {\textbf{Deterministic controller}\\ level $\cdot$ budget $\cdot$ retry $\cdot$ stop};

  \node[exec, text width=44mm] (ex) at (66,-50)
    {\textbf{Executor} (ReAct)\\ one Thought, one Action, one Observation};

  \node[box, fill=yellow!15, text width=42mm] (out) at (132,16)
    {\textbf{Answer} or \textbf{abstention}\,+\,failing level\\ + the evidence inspected};

  \draw[flow] (l1) -- (l2);
  \draw[flow] (l2) -- (l3);
  \draw[flow] (l3.south) -- (0,-50) -- (ex.west);
  \draw[flow] (ex.east) -- (132,-50) -- (r1.south);
  \draw[flow] (r1) -- (r2);
  \draw[flow] (r2) -- (r3);
  \draw[flow] (r3) -- (out);

  \draw[mediate] (l1.east) -- (41,0);
  \draw[mediate] (l2.east) -- (41,-17);
  \draw[mediate] (l3.east) -- (41,-34);
  \draw[mediate] (91,0) -- (r3.west);
  \draw[mediate] (91,-17) -- (r2.west);
  \draw[mediate] (91,-34) -- (r1.west);
  \draw[mediate] (66,-42) -- (ex.north);
  \node[mem, text width=54mm] (mem) at (66,17)
    {\textbf{Context store}: success ledger $\cdot$ failure index\\
     written only by verifier verdicts; read on assembly};
  \draw[mediate] (mem.south) -- (66,8);
\end{tikzpicture}}
\caption{The architecture. Dotted lines are the controller's mediation: no
artifact reaches a verifier and no verdict reaches a generator except through
it, and it alone reads and writes the context store (\S\ref{sec:memory}). Each
verifier is a free deterministic gate followed by an optional LLM judge invoked
only on what the gate cleared; a rejection routes to its paired level, and an
exhausted budget yields an abstention naming that level. The pilot implements
the R3 and R1 rows only.}
\Description{A block diagram. Specification levels L1 to L3 form a column on the left, verifiers R1 to R3 a column on the right, and a vertical deterministic controller sits between them. Dotted double-headed arrows connect every specification and verifier to the controller. An executor box sits below, and a context store above. Output at top right is an answer or an abstention naming the failing level.}
\label{fig:arch}
\end{figure*}

Figure~\ref{fig:vmodel} shows the motivating frame and Figure~\ref{fig:arch}
the architecture it yields: specification levels descend on the left, verifiers
ascend on the right, and each verifier is paired with exactly one level, so the
level at which a failure is caught identifies the level at which it was
introduced.

\emph{L1, requirements}: the request plus a machine-usable statement of what
would satisfy it, such as the expected answer type or, in code, the failing test
that must pass. \emph{L2, decomposition}: the subtasks, their order, the data flow between
them, and each one's completion condition; this is where most long-horizon
failures are introduced, since a decomposition that mis-binds one subtask to
another cannot be repaired downstream.
\emph{L3, step specification}: one tool, one argument, one expected result type;
the only level the executor sees in full.

\emph{R1, unit} inspects one observation against the L3 step that requested it.
\emph{R2, integration} inspects the assembled intermediate results against L2: do
references resolve, do types compose, do the parts concern the entities the
decomposition assumed. \emph{R3, acceptance} inspects the final answer against
L1.

\subsection{Runtime properties}
\label{sec:arch}

Three properties of the architecture are load-bearing. First, \emph{the
controller is the only stateful component}: it holds the current level, the
per-level budget, and the artifact store, and the generative components are
called as functions that return content. Second, \emph{every verifier is split}:
the gate half runs first and costs nothing, and the judge half runs only on
artifacts the gates have cleared, and only at levels whose failure class is
semantic; this split is what makes the two halves separately priceable. Third,
\emph{rejection routes to the paired level}: R1 re-issues one step; R2 returns
to L2 and re-decomposes, discarding only what is downstream of the mis-binding;
R3 returns to L1 or, once the budget is exhausted, abstains. The system's
output is therefore a triple: an answer or an abstention, the level of the last
rejection, and the evidence inspected.

One consequence holds by construction rather than by model behavior.
\emph{Bounded termination with cause}: budgets are finite constants and every
transition is decided by finite program logic, so every run terminates within
bounded steps in one of exactly two states---an accepted answer or an
abstention naming the failing level---under no assumption about the backbone;
the uncapped collapse of \S\ref{sec:accounting} is what its absence looks like.

\subsection{Memory: what each component may remember}
\label{sec:memory}

\begin{figure}[h]
\centering
\resizebox{0.76\columnwidth}{!}{%
\begin{tikzpicture}[x=1mm,y=1mm, every node/.append style={font=\scriptsize}]
  \node[ctrl, text width=46mm] (asm) at (44,0)
    {\textbf{Assembler} (controller)\\ builds the context to the role's schema};
  \node[box, text width=46mm] (ctx) at (44,-15)
    {\textbf{Context for ONE invocation}\\ objective $\cdot$ contract $\cdot$ checks
     $\cdot$ this component's failures};
  \node[exec, text width=34mm] (ag) at (44,-29) {\textbf{Agent} produces one artifact};
  \node[ver, text width=40mm] (vf) at (44,-41) {\textbf{Verifier}: gate, then judge};

  \node[spec, text width=34mm] (succ) at (14,-58)
    {\textbf{Success ledger}\\ accepted artifacts only};
  \node[ver, text width=34mm] (fail) at (74,-58)
    {\textbf{Failure index}\\ component $\times$ condition};

  \draw[flow] (asm) -- (ctx);
  \draw[flow] (ctx) -- (ag);
  \draw[flow] (ag) -- (vf);
  \draw[flow] (vf.south) -- (44,-50) -- node[above,font=\tiny]{accept} (14,-50) -- (succ.north);
  \draw[flow] (vf.south) -- (44,-50) -- node[above,font=\tiny]{reject} (74,-50) -- (fail.north);
  \draw[mediate] (succ.west) -- (-10,-58) -- (-10,0) -- (asm.west);
  \draw[mediate] (fail.east) -- (98,-58) -- (98,0) -- (asm.east);
\end{tikzpicture}}
\caption{The write path. A verdict is the only event that writes memory, so a
rejected attempt survives as its verdict and not as a transcript, and the assembler
reads back only what Table~\ref{tab:context} permits.}
\Description{A flow diagram. An assembler builds the context for one invocation, which goes to an agent, whose artifact goes to a verifier. The verdict branches: accept writes to a success ledger, reject writes to a failure index. Dotted arrows return from both stores to the assembler.}
\label{fig:memory}
\end{figure}
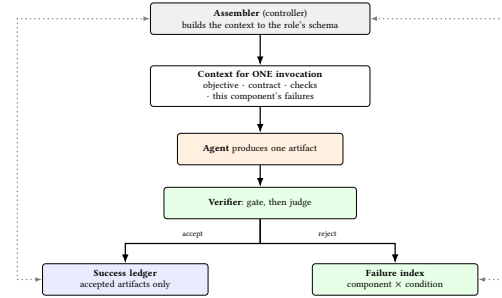

In a ReAct loop the transcript \emph{is} the memory: every thought, action,
observation, and failed attempt is resubmitted on each turn, which is why token
consumption grows quadratically~\cite{rewoo} and why, late in a task, most of
what the model reads concerns steps it has already finished or already got
wrong. Nothing in the loop decides what belongs there. Our design replaces
accumulation with assembly: the controller owns two stores, a verdict is the
only event that writes to either, and each invocation receives a context built
to a declared schema (Table~\ref{tab:context}, Figure~\ref{fig:memory}).

The \emph{success ledger} is append-only and written only when a verifier
accepts: the subtask, the condition it satisfied, the accepted output or
interface, and a pointer to the evidence; rejected attempts and raw transcripts
never enter it, so it is by construction the set of steps that moved the task
forward. The \emph{failure index} is keyed by component and condition,
recording which check rejected, why, and what the retry did; a retry receives
only the entries matching its own component and condition, which turns a pile
of failed attempts into a line reading \emph{rejected twice by the type gate
for the same missing field} and makes repetition a stopping condition the
controller can read. \emph{Isolation} makes both stores usable: the executor is
invoked per subtask, and a subtask is a new input rather than a continuation,
receiving its objective, contract, and checks---and none of the requirements,
architecture, sibling implementations, or earlier transcripts.

Two consequences follow, both design expectations rather than results: the
executor's context is bounded by the largest subtask instead of by trajectory
length, and abstention becomes \emph{scoped}, since work banked in the ledger
survives a component's failure. Neither is implemented in the pilot, which
keeps a conventional transcript with a truncation guard; \S\ref{sec:roadmap}
schedules the schema as an ablation.

\begin{table}[t]
\centering
\scriptsize
\setlength{\tabcolsep}{3pt}
\caption{The context schema. The executor row is what makes a subtask a new input
rather than a continuation.}
\label{tab:context}
\begin{tabular}{@{}p{0.18\columnwidth}p{0.37\columnwidth}p{0.35\columnwidth}@{}}
\toprule
\textbf{Role} & \textbf{Reads} & \textbf{Never reads} \\
\midrule
Planner (L2) & L1; inventory of tools and components & any execution output \\
Designer (L3) & L2; this component's slot and contract & sibling implementations \\
\textbf{Executor} (one subtask) & \textbf{its objective; the contract it must
honor; its checks; its own failure entries} &
\textbf{L1, L2, sibling code, every earlier transcript} \\
Gates & the artifact and the one condition & everything else \\
Judges R1--R3 & the artifact and its paired specification level & every other
level, and all transcripts \\
Controller & the whole state & --- \\
\bottomrule
\end{tabular}
\end{table}

\subsection{What is new, and what the pilot omits}
\label{sec:middle}

Each ingredient of the architecture has precedent; the arrangement does not.
Deterministic control of traversal appears in EoG~\cite{eog}, external per-step
signals appear in process supervision~\cite{verifystep} and interpreter
feedback~\cite{selfdebug}, and model-based critique appears in Reflexion and
Self-Refine~\cite{reflexion,selfrefine}. What is absent from all of them is a
verification subsystem that is \emph{stratified by specification level}, so that
rejections localize; \emph{external and binding} by construction, so that the
checked party cannot overrule the check; \emph{split into free and paid halves
at every level}, so that the cost of judging becomes an empirical question; and
coupled to a \emph{memory written only by verdicts}. O2 is the claim that this
arrangement, not any one part, is what turns an agent into a component other
software can call.

The pilot omits the middle (L2/R2) pair because multi-hop question answering
assembles nothing: every subtask is a retrieval returning a string, so an
integration verifier has nothing to inspect beyond what L3/R1 already compares.
This is a property of the task family, not of the architecture, and it is why
\S\ref{sec:roadmap} moves to code, where assembled artifacts and executable
ground truth both exist. \S\ref{sec:qaexample} presents a pilot failure that
only an integration verifier could have caught.

\subsection{Worked example: a four-hop question}
\label{sec:qaexample}

The item asks for the son of the Italian navigator who explored, for England, the
eastern coast of the continent on which Ignacio Esparza's birthplace lies; the gold
answer is \emph{Sebastian Cabot}. A VPG run produced this accepted plan (L2), with
what each subtask returned:

\noindent\begin{minipage}{\columnwidth}
\begin{quote}\footnotesize\setlength{\parskip}{0pt}
1. Find the region where Ignacio Esparza was born.\\
\hspace*{1em}$\rightarrow$ \emph{Guadalajara, Jalisco, Mexico} \quad(correct)\\
2. Find the navigator father of the person whose name is associated with the region
in the result of subtask 1.\\
\hspace*{1em}$\rightarrow$ \emph{Giovanni Battista Caviglia} \quad(wrong)\\
3. Find the explorer who explored the east coast of the region in the result of
subtask 2.\\
\hspace*{1em}$\rightarrow$ \emph{Lieutenant James Cook} \quad(wrong, inherited)\\
4. Find the navigator whose name is the result of subtask 3.\\
\hspace*{1em}$\rightarrow$ \emph{no answer; retrieval continues until the budget ends}
\end{quote}
\end{minipage}

Every R1 check passes at every step: each subtask returned a non-empty,
correctly typed result, free of tool errors and present in a retrieved
paragraph, and R3 never fires because no final answer is produced. The fault
lies at L2 and is invisible to both implemented pairs: subtask 3 is defined
over \emph{the result of subtask 2} rather than over the question, so the
distractor matched at step 2 propagates. This is an integration failure in the
V-model sense---individually valid units assembled into an invalid whole. An R2
gate comparing the type subtask 2 produced (a \emph{person}) with the type
subtask 3 presupposes (a \emph{region}) would reject the chain at step 3 with
no model call and return control to L2 for re-decomposition; the judge half
would handle the residual case in which types agree but entities do not.
Whether type-composition gates fire often enough to pay for themselves is an
empirical question the plan places first among the code-domain experiments.

\subsection{Worked example: fixing a repository issue}
\label{sec:swe}

A SWE-bench task supplies a repository, an issue, a test the patch must make
pass, and a suite that must keep passing~\cite{swebench}---two test sets at
different specification levels, precisely the structure the architecture
exploits. Consider an issue reporting that timestamps lose their timezone
offset:

\begin{itemize}\setlength{\itemsep}{0pt}\setlength{\parskip}{0pt}
\item[\emph{L1}] the issue plus its failing test.
\item[\emph{L2}] two files change, the serializer and the parser, and their
  contract is that a timestamp carries an offset.
\item[\emph{L3}] one edit at a time, each with the tests that cover it. The
  executor is given \emph{one edit, its contract and its tests}---not the
  issue, not the repository, not the previous edit's transcript.
\item[\emph{R1}] the patch applies, the file parses, its own tests pass. Free, and
  no model involved.
\item[\emph{R2}] the rest of the suite. The fix passes its own test and breaks
  three that parsed the old format: every edit is correct alone, the assembly is
  not. Control returns to L2, not to the executor, and only the edits downstream of
  the revised contract are redone.
\item[\emph{R3}] the original failing test, plus the one question no gate can
  answer: does the patch fix the issue, or only the test that came with it?
\end{itemize}

If an edit is rejected three times by the same gate, the run stops with
\emph{failed at R1, edit 2, patch does not apply}, keeping what was already
accepted. A monolithic agent emits patches until its budget ends and returns a
transcript in which nothing separates the useful attempt from the discarded ones.

\subsection{The pilot implementation}
\label{sec:pilot}

\begin{figure*}[t]
\centering
\resizebox{0.70\textwidth}{!}{%
\begin{tikzpicture}[node distance=9mm, every node/.append style={font=\scriptsize}]
  \node[box] (q) {Question};
  \node[spec, right=of q] (pl) {Planner\\(LLM)};
  \node[gate, right=of pl] (pg) {Plan\\gates};
  \node[ver, right=of pg] (pj) {Plan judge\\(LLM)};
  \node[exec, right=of pj] (ex) {Executor loop\\Thought\,$\to$\,Action\\$\to$\,Observation};
  \node[gate, right=of ex] (ag) {Answer\\gates};
  \node[ver, right=of ag] (aj) {Answer judge\\(LLM)};
  \node[box, right=of aj, fill=yellow!15] (out) {Answer\\or abstain};

  \draw[flow] (q) -- (pl);
  \draw[flow] (pl) -- (pg);
  \draw[flow] (pg) -- node[above,font=\tiny]{pass} (pj);
  \draw[flow] (pj) -- node[above,font=\tiny]{pass} (ex);
  \draw[flow] (ex) -- (ag);
  \draw[flow] (ag) -- node[above,font=\tiny]{pass} (aj);
  \draw[flow] (aj) -- node[above,font=\tiny]{pass} (out);

  \draw[back] (pg.north) .. controls +(0,6mm) and +(0,6mm) ..
      node[above,font=\tiny]{reject: regenerate plan} (pl.north);
  \draw[back] (pj.north) .. controls +(0,9mm) and +(0,9mm) ..
      node[above,font=\tiny]{reject} (pl.north east);
  \draw[back] (ag.south) .. controls +(0,-6mm) and +(0,-6mm) ..
      node[below,font=\tiny]{reject: retry step} (ex.south);
  \draw[back] (aj.south) .. controls +(0,-12mm) and +(0,-12mm) ..
      node[below,font=\tiny,yshift=-1mm]{reject} (ex.south east);

  \node[ctrl, below=13mm of ex, minimum width=110mm]
     {\textbf{Deterministic controller}: owns every transition, budget, retry and stop.
      Agents propose content only.};
\end{tikzpicture}}
\caption{The implemented pipeline. Diamonds are deterministic checks (no model
cost); green rounded boxes are LLM judges, invoked only on gate-cleared
artifacts; dashed arrows are bounded rejection paths. When plan regeneration is
exhausted, the system falls back to executing the question as a single task
(\S\ref{sec:results}).}
\Description{A left-to-right pipeline. A question passes through a planner, plan gates, a plan judge, an executor loop, answer gates and an answer judge to an answer or abstention. Dashed arrows loop backwards from each check to the component it rejects. A deterministic controller spans the width beneath.}
\label{fig:workflow}
\end{figure*}
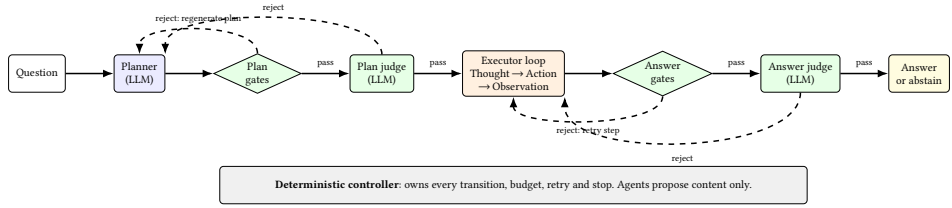

The implemented pipeline (Figure~\ref{fig:workflow}) is Figure~\ref{fig:arch}
without the R2 row and without the context store. Three principles govern the
implementation. First, control resides in program logic: steps, retries, and
termination are decided in ordinary code bounded by named constants. Second,
inexpensive checks precede expensive ones, so that a judge is paid for only
where gates cannot decide. Third, failures are reported rather than
reinterpreted, because silently reformulating a step into a weaker objective is
the mechanism by which agents come to fabricate values.

Eleven gates examine the plan: nine structural checks (parseability, length, duplicate subtasks, self-, forward and dangling references, terminal step, typed completion conditions, tool existence) and two that apply only to arithmetic questions (single-expression form and numeric coverage). Seven examine an answer (presence, non-emptiness, non-answers, consistency with the alternatives a comparison offers, tool errors, type agreement, and optionally occurrence in an observation). The plan judge asks whether the steps,
assuming each succeeds, would answer the question; the answer judge asks whether the
evidence entails the answer. Three guards operate inside the executor---a
repeated action is served from cache with a warning, unparseable output ends
the run after bounded attempts, and observations are truncated before entering
the transcript; per \S\ref{sec:defn} these are loop control, not verification.

\begin{table}[h]
\centering
\footnotesize
\setlength{\tabcolsep}{3.5pt}
\caption{The five configurations; internal labels in parentheses match the released
logs. All share one executor, tool set, retriever, scorer and backbone, so adjacent
pairs differ by one component.}
\label{tab:arms}
\begin{tabular*}{\columnwidth}{@{\extracolsep{\fill}}llcccc@{}}
\toprule
\textbf{Config.} & \textbf{Composition} & \textbf{Plan} & \textbf{Guards} & \textbf{Gates} & \textbf{Judges} \\
\midrule
N (A)    & plain ReAct              & --  & --  & --  & --  \\
NP (B)   & ReAct + planner          & yes & --  & --  & --  \\
VG (C1)  & verified, no planner     & --  & yes & yes & --  \\
VPG (C0) & + planner + plan gates   & yes & yes & yes & --  \\
VPGJ (C) & + LLM judges             & yes & yes & yes & yes \\
\bottomrule
\end{tabular*}
\end{table}

Table~\ref{tab:arms} defines the configurations. VG proves central to the
analysis: it has no planner---the question itself is the single task, carrying a
completion condition derived from its interrogative form, so that ``Who \ldots''
yields \emph{a person or organization, not one of their attributes}. VG is the
minimal configuration that retains verification.

\section{Experimental Design}
\label{sec:setup}

Establishing O2 requires five methodological ingredients, most of which the
pilot lacks by design: (i)~\emph{assembled artifacts}, without which R2 has
nothing to inspect; (ii)~\emph{executable ground truth}, without which a
verifier's mistakes cannot be distinguished from the scorer's;
(iii)~\emph{several model tiers}, because each half of the layer addresses a
failure class that appears at a different scale; (iv)~\emph{abstention-aware
measures}, because accuracy alone cannot summarize a system that is permitted to
decline; and (v)~a \emph{switchable memory schema}, without which context
management and verification are measured together. The plan in
\S\ref{sec:roadmap} is organized around supplying all five. The pilot has the
fourth and part of the third: it is a feasibility study, showing that
architectural differences are measurable at all when every other component is
held constant.

\paragraph{Task, tools, model.}
MuSiQue-Answerable~\cite{musique}, four-hop stratum: twenty paragraphs per item, of
which roughly four support the answer and the rest are mined distractors. We use
five items, one of which also appears in a four-paragraph gold-context variant as a
control of reduced difficulty. Two tools are bound to a single item's paragraphs:
\texttt{retrieve[query]}, a BM25~\cite{bm25} index returning three query-centred
excerpts with indices, and \texttt{read\_\allowbreak paragraph[idx]}. A single
allow-list function is the only path from harness to tool, emitting nothing
beyond index, title, and text, so the gold answer and the supporting-paragraph
annotations cannot reach the retriever. Every role in
every configuration runs \texttt{llama-3.1-8b-instant} at temperature 0, so
differences are attributable to architecture and not capacity; the executor budget
is 12 steps unless stated otherwise, over two passes.

\paragraph{Measures.}
A verified configuration may decline to answer and an unverified one cannot,
which a single accuracy figure cannot represent. We therefore report a partition
summing to unity: \emph{correct}; \emph{silent error} (an answer was produced
and is wrong---the consequential case, since nothing downstream can distinguish
it from a correct answer); and \emph{abstention}; plus \emph{coverage}, the
proportion of scoreable runs that answered. Scoring follows MuSiQue's
SQuAD-style procedure~\cite{squad}. A \emph{correction} is a rejection whose
retry produced a \emph{different} answer that was then accepted; the change
requirement matters, since accepting a resubmitted identical answer is the
judge contradicting itself.

\paragraph{Execution accounting.}
\label{sec:accounting}
The study comprises \emph{47 executions}: five items over two passes for four
configurations, plus seven for VPGJ, whose second pass was replaced by two runs
with the plan judge disabled to observe an executed plan.
Of these, 37 were scoreable and 10 were not, a distribution that is itself a
result (Table~\ref{tab:accounting}). Three points follow. First, the
second pass for N and NP ran with the step cap removed, and \emph{all ten of
those runs terminated without scoreable output}, each invoking the same tool
until the provider's token limit ended the process. Second, repeated passes for
VG and VPG reproduced the first pass exactly---decoding determinism at
temperature 0, not stability---so the per-run costs in Table~\ref{tab:cost} are
computed over distinct logged traces. Third, the plan judge in VPGJ rejected every plan in all
five of its enabled runs, and the two runs with the judge disabled both returned
no answer; this is the smallest cell in the study and is reported separately.

\section{Results of the Pilot}
\label{sec:results}

\subsection{Execution outcomes (RQ1)}

\begin{table}[h]
\centering
\footnotesize
\setlength{\tabcolsep}{3pt}
\caption{Execution accounting over all 47 runs. No configuration produced a silent
error, so that column is omitted.}
\label{tab:accounting}
\begin{tabular*}{\columnwidth}{@{\extracolsep{\fill}}lrrrrrr@{}}
\toprule
\textbf{Config.} & \textbf{exec.} & \textbf{score.} & \textbf{crash} &
\textbf{corr.} & \textbf{abst.} & \textbf{cov.} \\
\midrule
N (plain ReAct)   & 10 & \phantom{0}5 & \textbf{5} & \textbf{0} & 5 & 0.00 \\
NP (+ planner)    & 10 & \phantom{0}5 & \textbf{5} & \textbf{0} & 5 & 0.00 \\
VG (verified)     & 10 & 10 & 0 & \textbf{8} & 2 & 0.80 \\
VPG (+ planner)   & 10 & 10 & 0 & 1 & 9 & 0.10 \\
VPGJ (+ judges)   & \phantom{0}7 & \phantom{0}7 & 0 & \textbf{4} & 3 & 0.57 \\
\midrule
\textbf{Total}    & \textbf{47} & \textbf{37} & \textbf{10} & \textbf{13} & \textbf{24} & --- \\
\bottomrule
\end{tabular*}
\end{table}

N and NP answered none of the ten questions: all ten scoreable runs hit the
12-step cap without once invoking \texttt{finish}, all ten uncapped runs
continued until the provider refused them, and the study's entire 21\% crash
rate falls on these two configurations. VG answered eight of ten,
using fewer steps than N (Table~\ref{tab:cost}), and abstained on the two items
it could not resolve instead of guessing.

\subsection{A confound between guards and gates}
\label{sec:confound}

One qualification applies to every result below. The verified configurations
differ from N and NP in two respects: the gates, which are verification, and the
three guards, which are loop control. The collapse mode of
\S\ref{sec:noabstain}, in which an agent invokes the same tool until an external
limit intervenes, is plausibly prevented by the duplicate-action guard alone,
which refuses the second identical invocation without consulting any gate. We
therefore cannot attribute the observed difference to verification as such;
separating the two is the first ablation in \S\ref{sec:roadmap}.

\subsection{The effect of planning (RQ3)}
\label{sec:plan}

VPGJ with the plan judge enabled and VG produced identical outcomes: the judge
rejected all five plans, so VPGJ fell back to executing the question as a single
task and became VG with additional overhead. Partitioning every run by whether a
plan was executed makes the effect explicit. Twelve runs executed a plan, of
which one was correct (a success rate of 0.08), at a mean of 15.6 executor
steps; fifteen runs executed no plan, of which twelve were correct (0.80), at a
mean of 10.9 steps. Planning thus cost 43\% more steps for roughly one tenth of
the success rate.

Two mechanisms appear in the traces. \emph{Budget fragmentation}: four subtasks
under a twelve-step budget allow three steps each, the minimum viable trajectory
(retrieve, read, finish), leaving no capacity to reformulate a failed query, and
most planned failures end at the first or second subtask. \emph{Error propagation}:
the chain fails at its first incorrect resolution, as \S\ref{sec:qaexample} shows in
full, and no gate in the implemented system can detect it, since each subtask answer
is present, non-empty, typed and grounded and only the composition is wrong; no
configuration solved that item in any run. This is the missing L2/R2 pair
appearing as a measured result rather than an argument.

\subsection{Cost (RQ4)}
\label{sec:cost}

\begin{table}[t]
\centering
\footnotesize
\setlength{\tabcolsep}{4pt}
\caption{Mean cost per scoreable run over distinct logged traces. Executor steps are
the architecture-neutral comparison; total calls include planner and judge overhead.}
\label{tab:cost}
\begin{tabular*}{\columnwidth}{@{\extracolsep{\fill}}lrrrr@{}}
\toprule
\textbf{Config.} & \textbf{LLM calls} & \textbf{exec.\ steps} &
\textbf{prompt tok.} & \textbf{compl.\ tok.} \\
\midrule
N     & 12.0 & 12.0 (all at cap) & 18{,}891 & \phantom{1,}598 \\
NP    & 13.0 & 12.0 (all at cap) & 11{,}180 & \phantom{1,}752 \\
VG    & \textbf{11.8} & \textbf{11.8} (3--21) & 19{,}511 & 1{,}150 \\
VPG   & 15.2 & 14.2 (8--20) & 17{,}226 & \phantom{1,}840 \\
VPGJ  & 14.8 & 10.0 (4--18) & 20{,}943 & 1{,}210 \\
\quad{\itshape judge disabled} & 23.0 & 20.0 & 28{,}609 & 1{,}464 \\
\bottomrule
\end{tabular*}
\end{table}

The distribution matters more than the mean: N and NP sat at the cap in every
scoreable run, having no way to stop earlier, whereas VG ranged from 3 to 21
steps---resolving one item in three and spending its full allocation plus a
retry on the item it declined---the observable signature of adaptive
termination.

\subsection{Attribution of corrections (RQ2)}

\begin{table}[t]
\centering
\footnotesize
\setlength{\tabcolsep}{4pt}
\caption{Rejections and corrections across the logged traces; identical repeats are
not double-counted. A correction required the retry to produce a different answer
that was then accepted.}
\label{tab:attrib}
\begin{tabular*}{\columnwidth}{@{\extracolsep{\fill}}lrrrl@{}}
\toprule
\textbf{Source} & \textbf{rejections} & \textbf{corrections} & \textbf{rate} & \textbf{marginal cost} \\
\midrule
Deterministic gates & 24 & \textbf{8} & 0.33 & none \\
LLM judges          & \phantom{0}3 & \phantom{0}1 & 0.33 & one call each \\
\bottomrule
\end{tabular*}
\end{table}

The two halves of the layer show a comparable success rate per rejection but
differ by an order of magnitude in volume: gates fired 24 times, judges three
times, with no judge reversal (rejecting and then accepting the identical
answer). The plan judge behaved differently from the answer judge: it
rejected all five plans with a recurring objection of the form \emph{``the plan
does not guarantee that $X$ is the same as $Y$''}, twice identically across both
regeneration attempts. No plan can guarantee what a retrieval will return, so
regeneration cannot satisfy such an objection, and the fallback to unplanned
execution becomes deterministic.

\subsection{Retrieval ceiling}
\label{sec:ceiling}

The agent reached 67\% of the required paragraphs on average, and all of them in
25\% of runs, which bounds every configuration: architectural differences are
measured on a retriever that omits about a third of the necessary evidence,
because the executor issues sentence-form queries against a lexical index. VG
answered eight of ten while reaching complete evidence in only one run of five,
so several items were resolved from a subset of the annotated evidence.

This ceiling bounds verifier calibration in both directions. The judge never
sees the gold answer or the supporting-paragraph annotations, so when required
evidence is never fetched it can reject an answer the scorer would have
accepted, inflating abstentions. We observed no such false negative: the three
judge rejections concerned intermediate results the scorer does not evaluate,
and every final-answer abstention came from a gate observing that no answer
existed. We observed no false positives either, but attach little weight to
that, since no configuration produced a confidently wrong answer for a judge to
accept.

\section{Discussion}
\label{sec:discussion}

\subsection{Unverified agents do not abstain}
\label{sec:noabstain}

No unverified execution ended with a refusal: the recorded abstentions of N and
NP in Table~\ref{tab:accounting} are the harness observing that the loop ended
without invoking \texttt{finish} once the step cap intervened, and removing the
cap---our artifact, not ReAct's---produced the uncapped collapse of
\S\ref{sec:accounting}. The verified configurations differ in kind: every
termination is a decision with a recorded cause. Independently of accuracy,
halting with a stated reason is what lets a component be composed into a larger
system; we therefore regard the deterministic half as close to a precondition
for unattended deployment, subject to the confound of \S\ref{sec:confound}.

\subsection{Behavior under a stronger backbone}
\label{sec:stronger}

No confidently wrong answers occurred, so the silent-error hypothesis could not
be tested; we report this as a null result. Using the model on which the judge
contributes least was deliberate: \emph{the 8B experiment establishes a lower
bound}. When reasoning capacity is identical and low across configurations,
whatever separates them is attributable to guards, gates, and control flow; at
70B the backbone absorbs the failures the architecture exists to handle,
measuring the model instead of the design. The cost is that the judge's target
failure class does not arise.

Exploratory runs on two larger backbones
(\texttt{llama-3.3-70b-\allowbreak versatile} and
\texttt{qwen/\allowbreak qwen3.6-27b}), outside the controlled set and
requiring confirmation, indicate where that failure class does arise. With a
stronger backbone, N and NP stop collapsing structurally: they hold the action
format, terminate, and produce answers. On harder items, however, the loop
commits to a plausible intermediate entity and returns a fluent answer with no
indication that the chain broke. A recurring proportion of these answers were
rejected by the answer judge and by no gate, for reasons citing exactly what
gates cannot evaluate: the evidence does not support the answer, the
observation concerns a different entity, or a plausible name was inferred
rather than retrieved---the intrinsically semantic class the hallucination
literature~\cite{hallucination,selfcheckgpt} targets. The implication is a
division of labor: gates are mandatory, being free and covering the malformed,
empty, mistyped, repeated, and placeholder outputs that dominate at every scale
we examined; judging should be deployed selectively, where unsupported but
well-formed output is the anticipated failure.

\subsection{Planning and verification are partial substitutes}

Also outside the controlled set, on stronger backbones NP consistently
outperformed N, in line with ReWOO and ADaPT~\cite{rewoo,adapt}; that advantage
did not survive verification (8 of 10 versus 1 of 10). A tentative explanation
is that a planner's benefit to an unverified loop is structural discipline,
which guards and gates supply more cheaply; what remains is the decomposition
itself, which on retrieval-shaped tasks fragments the budget and propagates the
first error. The architectural reading is that a decomposition is an L2
artifact and the pilot has no R2 verifier to check one; degradation from
adding L2 without R2 is what the architecture predicts. Whether planning recovers its value once
paired with integration verification is a question for \S\ref{sec:roadmap}.

\subsection{Why self-assessment is structurally limited}
\label{sec:selfassess}

We attribute the judge's weakness at 8B to its configuration, not to prompt
quality; the prompts were revised repeatedly, and the final failures used the
revised rubric's own vocabulary. The judges are the same model at the same
temperature, examining the same tool outputs and evaluating the same model's
work, so they share the actor's blind spots and are least reliable exactly where
the actor is. Their available behaviors reduce to agreement, which conveys
nothing, and unanchored disagreement, which is noise. This is the configuration
Huang et al.~\cite{selfcorrect} report to fail, and it violates V1 in spirit
while satisfying it in letter.

Two designs would constitute a fair test of model-based judging.
\emph{Asymmetric verification} places a strictly stronger model in the judge
roles above a weaker executor; judges issue one to two calls per subtask
against the executor's seven to thirteen, so the cost is modest
(\S\ref{sec:stronger} provides indirect evidence in its favor).
\emph{Ground-truth-constrained verification} allows the judge only questions
whose evidence the deterministic layer supplies (e.g., whether a specified span
entails the answer~\cite{verifystep,selfdebug}); its limit is that it cannot
detect a failure the deterministic layer did not already suspect.

Finally, what the judge did catch---an executor emitting the literal
placeholder \texttt{finish[\ldots]}, admitted by all seven gates---was
syntactic, and expressing it as a gate afterwards took three lines of code.
This suggests a practice: run a judge during construction, promote each
recurring catch into a gate, and reserve run-time judging for the semantic
class above, which cannot be promoted.

\section{Threats to Validity}
\label{sec:threats}

We follow the threat categories conventional in empirical software
engineering~\cite{wohlin,runeson}.

\emph{Internal validity.} The five configurations are implemented as separate
directories; the executor, gates, tools, and scorer are byte-identical across
them and every run's configuration is logged, but divergence remains a risk we
detected and corrected more than once during development.

\emph{Construct validity.} Exact-match scoring under-credits paraphrases, and
the retrieval ceiling (\S\ref{sec:ceiling}) bounds every reported figure and may
generate false negatives for the judge.

\emph{External validity.} The study uses one backbone, one benchmark, one task
family, and the small sample of \S\ref{sec:accounting}; \S\ref{sec:stronger}
reports qualitative patterns, not measurements; and the architecture is
evaluated only in part, the integration row and the memory schema being argued
for rather than tested.

\emph{Conclusion validity.} With 37 scoreable runs and cells of 7 to 10, no
statistical inference is performed and none should be drawn; the absent runs
are not missing at random but are the crashes of Table~\ref{tab:accounting}, so
N and NP appear in the outcome columns only in their capped configuration.

\emph{Researcher bias.} One author implemented all five configurations, and
scoring was not blind.

\section{One-Year Research Plan}
\label{sec:roadmap}

Four deficiencies drive the plan: insufficient statistical power, a single
model tier, a task family that cannot exercise the integration layer, and
separate per-configuration directories. Table~\ref{tab:roadmap} organizes the
schedule around them.

\begin{table}[H]
\centering
\scriptsize
\linespread{0.92}\selectfont
\setlength{\tabcolsep}{3pt}
\caption{Twelve-month plan; half the schedule goes to code, where deterministic
verifiers are rich and the L2/R2 row can be exercised.}
\label{tab:roadmap}
\begin{tabular}{@{}p{0.1\columnwidth}p{0.855\columnwidth}@{}}
\toprule
\textbf{Months} & \textbf{\qquad Work and rationale} \\
\midrule
1--2 &
\textbf{Consolidation, and the confound.} One codebase with the configuration
as a parameter, removing the divergence risk of \S\ref{sec:threats}; a dense
retriever~\cite{dpr} to raise the 67\% ceiling. \emph{First ablation}: guards
without gates, gates without guards, and both ends of the ladder---until those
cells exist, the difference in \S\ref{sec:results} cannot be assigned to
verification rather than loop control. \\
\midrule
3--4 &
\textbf{Scale, and verifier configurations.} Roughly 500 items stratified by
hop count, three seeds, three to five model tiers, every repeat retained;
paired per-item analysis with effect sizes~\cite{dror}, hypotheses registered
beforehand. Asymmetric and ground-truth-constrained verification
(\S\ref{sec:selfassess}) as separate conditions. \emph{Second ablation}: full
accumulation vs.\ ledger only vs.\ both stores, with tokens read per step
reported alongside accuracy. \\
\midrule
5--7 &
\textbf{Transfer to code.} Port to SWE-bench~\cite{swebench} and its agent
interfaces~\cite{sweagent}, with function-level suites under rigorous test
augmentation~\cite{humaneval,evalplus} as a cheaper inner loop. In code the
gate layer becomes a compiler, type checker, and test runner, with ground truth
no question-answering gate can have; we expect the gate/judge balance to shift
and will re-run Table~\ref{tab:attrib}. \\
\midrule
8--10 &
\textbf{Debugging, and the integration layer.} Fault localization and repair,
where a failing test supplies both trigger and oracle~\cite{selfdebug}. The
L2/R2 pair becomes testable at last: individually correct units forming a
broken assembly is a real repository failure class invisible to unit checks
(\S\ref{sec:qaexample}). Repository length makes the memory schema
consequential, so its ablation is repeated with per-component isolation on and
off. \\
\midrule
11--12 &
\textbf{Synthesis.} Cross-domain comparison of where each half contributes, a
replication package of code, prompts, items, seeds and logs, and submission. \\
\bottomrule
\end{tabular}
\end{table}

\section{Conclusion}

This paper argued that LLM agents run a delivery process without the
certification structure such a process needs, transferred the V-model to
supply that structure, and measured its deterministic core. The measurement
is small---five items, 47 executions, one 8B backbone---but its central
contrast is stark: without the layer, no run ever terminated by decision;
with it, every run ended in an accepted answer or a reasoned abstention, and
eight of nine corrections were free. Planning, added without its paired
verifier, made the system worse---the architecture's own prediction,
appearing as data.

Three uncertainties bound what this establishes, and they are the feedback we
seek at the workshop: whether the effect belongs to verification or to the
bundled guards (\S\ref{sec:confound}); whether judges recover their value
once separated from the model they check (\S\ref{sec:selfassess}); and
whether the integration level, which no question-answering task can exercise,
earns its place once the architecture moves to code (\S\ref{sec:roadmap}).
What we ask the field to weigh, meanwhile, is not another technique for a
leaderboard, but whether an agent without an external verification layer is
structurally incomplete, in the way untested software is.

\balance
{\footnotesize
\setlength{\itemsep}{0pt}

}

\end{document}